\documentclass[aps,prb,superscriptaddress,twocolumn,showpacs]{revtex4}
\usepackage{graphicx}
\usepackage{amsmath}

\begin{document}

\title{Photoluminescence spectroscopy of trions in quantum dots: a theoretical description}

\author{Juan I. Climente}
\email{climente@unimo.it}
\affiliation{CNR-INFM National Center on nanoStructures and bioSystems at Surfaces (S3),
Via Campi 213/A, 41100 Modena, Italy}
\affiliation{Departament de Qu\'{\i}mica F\'{\i}sica i Anal\'{\i}tica, Universitat Jaume I,
Avda. Sos Baynat s/n, 12080 Castello, Spain}
\author{Andrea Bertoni}
\affiliation{CNR-INFM National Center on nanoStructures and bioSystems at Surfaces (S3),
Via Campi 213/A, 41100 Modena, Italy}
\author{Guido Goldoni}
\affiliation{CNR-INFM National Center on nanoStructures and bioSystems at Surfaces (S3),
Via Campi 213/A, 41100 Modena, Italy}
\affiliation{Dipartimento di Fisica, Universit\`a di Modena e Reggio Emilia,
Via Campi 213/A, 41100 Modena, Italy}
\date{\today}

\begin{abstract}
We present a full configuration interaction study of the spontaneous recombination
of neutral and singly charged excitons (trions) in semiconductor quantum dots, 
from weak to strong coupling regimes. 
We find that the enhancement of the recombination rate of neutral excitons with increasing 
dot size is suppressed for negative trions, and even reversed for positive trions. 
Our findings agree with recent comprehensive photoluminescence experiments in self-assembled quantum dots 
[P. Dalgarno et al.  Phys. Rev. B {\bf 77}, 245311 (2008)] and confirm the major role played by correlations 
in the valence band. 
The effect of the temperature on the photoluminescence spectrum and that of the
ratio between the electron and hole wavefunction lengthscales are also described.

\end{abstract}


\maketitle

\section{Introduction}

Optoelectronics is one of the fields where the atomic-like properties of quantum dots (QDs)
is liable of producing a breakthrough in device performance.\cite{Grundmann_book}
Novel quantum-optical applications, such as single-photon sources\cite{ZwillerNJP}
and quantum logic gates\cite{KrennerNJP,ChenSCI,BiolattiPRL,TroianiPRBPRL,KirazPRA,StinaffSCI}
have been proposed and are currently under intense research.
Progress in this research field has been made possible by a number of studies over the last
years which provide understanding of exciton and multiexciton emission processes taking place in QDs.\cite{Grundmann_book,Pawel_book}
Many of these studies highlight the critical effect of Coulomb interactions among
the particles confined in the QD to determine the photoluminescence (PL) response.
Indeed, it has been shown that the relative spectral positions of exciton ($X^0$)
and charged exciton ($X^{\pm n}$) states can be generally inferred from the few-particle
correlations.\cite{LelongSSC,HohenesterPSSaPSSb,WilliamsonEPL,ShumwayPRB,BesterPRB,AliPE,
HartmannPRL,EdigerNP,RegelmanPRB,FinleyPRB,NarvaezPRB,LandinSCIPRB,LomascoloPRB,RodtPRB}

The dependence of the recombination rates of neutral exciton, bi-exciton and multi-exciton complexes
on Coulomb interactions has been also described in a number of papers.\cite{RinaldiPRB,BryantPRB,QuePRB,CorniPRB,WimmerPRB,BacherPRL,SantoriPRB}
Much less is known about the recombination rates of positively ($X^+$) and negatively ($X^-$)
charged excitons (trions).\cite{NarvaezPRB,DalgarnoPRB,ChengPRB}
This is nonetheless an interesting problem.
From a fundamental point of view, comparing the recombination of $X^0$, $X^+$ and $X^-$
provides direct insight into the different role played by electrons and holes as exciton components.
From a practical point of view, trions are often the dominant species in QDs populated
by exciting valence electrons into the conduction band continuum.\cite{LomascoloPRB}
Moreover, they are involved in relevant applications of QDs
such as the optical preparation of pure spin electrons\cite{AtatureSCI} or
holes\cite{GerardotNAT,EbbensPRB}. They are also the natural source of single photons
with pure circular polarization, since neutral excitons are subject to
electron-hole exchange interaction and hence produce linearly polarized photons.
Understanding the conditions which optimize the emission of trions is of primary importance
for the development of such applications.

Very recently, Dalgarno and co-workers collected comprehensive spectroscopical data
comparing the recombination rates and energies of neutral excitons and trions in
self-assembled InGaAs QDs.\cite{DalgarnoPRB}
Clear differences were observed upon charging $X^0$ with an additional electron or hole,
which were qualitatively consistent with a picture where electrons are in the strong confinement
regime while holes are not. Owing to their larger mass, holes seemed to be in an intermediate
confinement regime where the influence of Coulomb correlations may be significant.\cite{EdigerNP}

In this work, we provide theoretical assessment on this subject. We
perform a full configuration interaction (FCI) study to calculate the
recombination rates of $X^0$, $X^+$ and $X^-$ in QDs with different confinement strength.
With increasing QD size, $X^0$, $X^+$ and $X^-$ display different trends that can be
traced back to the interplay among the different Coulomb interaction terms in the system
(electron-electron, hole-hole, electron-hole). In particular, we obtain an enhanced
PL for $X^0$ with increasing dot size -- a result well-known from previous
studies\cite{BryantPRB,QuePRB,CorniPRB}-- but then show that electron-electron repulsion partially
quenches the emission efficiency for $X^-$. For $X^+$ the quenching is even larger, due to the 
strong hole-hole correlation, and the reverse trend (decreasing PL with increasing dot size) is obtained. 
Indeed, holes are found to be significantly correlated even in typical self-assembled structures, 
which leads to spectral features in agreement with Ref.~\onlinecite{DalgarnoPRB}.
We also investigate the effect of the different lateral extension of single-particle electron and
hole wavefunctions on the PL spectrum of trions, and show that this electron-hole asymmetry 
may bring about qualitative changes in both the relative spectral positions and the relative recombination 
rates of $X^0$, $X^+$ and $X^-$.
Finally, we predict that hole-hole correlations may be reflected in a peculiar temperature dependence of 
the $X^+$ line, which does not show up for neutral and negatively charged excitons.

\section{Theory}

We use a Hamiltonian describing $N_e$ electrons and $N_h$ heavy holes confined in a QD,

\begin{equation}
{\cal H} = {\cal H}_e^{N_e} + {\cal H}_h^{N_h}  + V_{e h}^{N_e N_h}.
\label{eq1}
\end{equation}

\noindent
Here ${\cal H}_e^{N_e}$ and ${\cal H}_h^{N_h}$ are the electron and hole effective mass
Hamiltonians, respectively, including intra-band interactions

\begin{subequations}
\begin{align}
{\cal H}_e^{N_e} &= \sum_{i=1}^{N_e} \left[ \frac{\mathbf{p}_i^2}{2 m_e^*} + V_e(\rho_i,z_i) +
\sum_{j<i} \frac{e^2}{\epsilon |\mathbf{r}_{e}^{i}-\mathbf{r}_{e}^{j}|} \right],\\
{\cal H}_h^{N_h} &= \sum_{i=1}^{N_h} \left[ \frac{\mathbf{p}_i^2}{2 m_h^*} + V_h(\rho_i,z_i) +
\sum_{j<i} \frac{e^2}{\epsilon |\mathbf{r}_{h}^{i}-\mathbf{r}_{h}^{j}|} \right],
\end{align}
\end{subequations}
while $V_{e h}^{N_e N_h}$ is the Coulomb interaction term between
electrons and holes,

\begin{equation}
V_{e h}^{N_e N_h} \,=\, - \sum_{i=1}^{N_e} \sum_{j=1}^{N_h}
\, \frac{e^2}{\epsilon |\mathbf{r}_{e}^{i}-\mathbf{r}_{h}^{j}|}.
\label{eq2}
\end{equation}

\noindent
In the above expression, $m^*_\alpha$ and $V_\alpha(\rho,z)$ are the effective mass and the 
single-particle QD confinement potential, respectively, which is in general different for the two types of carriers.
$\epsilon$ the dielectric constant of the QD medium and $e$ the electron charge. We neglect the 
electron-hole exchange interaction. This does not affect
the singlet states of the trions we will deal with, and adds only a fine structure
to the exciton spectrum which is not relevant for our study.\cite{BayerPRB,BesterPRB}

We consider a separable confining potential with cylindrical symmetry, $V(\rho,z)=V(\rho)+V(z)$.
In the in-plane direction we take a parabolic confinement, $V(\rho)=1/2(m^*\omega^2 \rho^2)$,
with $\omega$ as the characteristic frequency. This yields Fock-Darwin (FD) single-particle
states for electrons and holes\cite{Pawel_book}, which provide a transparent yet fairly
accurate starting point to describe the optics of QDs with different confinement regimes,
from weakly confined (etched)\cite{KalliakosAPL} to strongly confined (self-assembled)\cite{WojsPRB,RaymondPRL}.
In the vertical direction, the potential $V(z)$ is defined by a rectangular quantum well
provided by the band-offset between the QD and barrier materials.
The quantum well eigenstates are derived numerically.

The $N_e$-electron and $N_h$-hole states are calculated independently using a two-step FCI
approach in the basis of the FD single-particle states. 
First, we diagonalize ${\cal H}_e^{N_e}$ and ${\cal H}_h^{N_h}$ exactly, 
following the FCI method described in Ref.~\onlinecite{RontaniJPC}. 
The resulting few-electron ($\Psi^{N_e}$) and few-hole ($\Psi^{N_h}$) states contain an exact 
description of the intra-band correlations.
Second, Hartree products of the correlated electron and hole states obtained in the first step are 
used to represent the electron-hole term $V_{eh}^{N_e N_h}$, which is diagonalized exactly to 
describe inter-band correlations. 
This method provides fully correlated excitonic states ($X^{N_e N_h}$) which are needed for 
an accurate estimate of the recombination probability\cite{CorniPRB,WilliamsonEPL,WimmerPRB} and 
energy\cite{LelongSSC,HohenesterPSSaPSSb,WilliamsonEPL,ShumwayPRB,BesterPRB,AliPE,EdigerNP,RegelmanPRB,FinleyPRB}.
Moreover, it gives direct estimates of $V_{eh}^{N_e N_h}$, which allows us to study the
renormalization of the electron-hole attraction upon charging with additional electrons
or holes.

The PL spectra are calculated within the dipole approximation and Fermi's golden rule.\cite{Pawel_book}
The recombination probability from an initial state $| X^{N_e N_h}, i \rangle$
to a final state $|X^{N_e-1 N_h-1}, f \rangle$ with one less electron-hole pair, at an emission frequency 
$\omega$ is then given by:

\begin{multline}
{\cal P}_{f \leftarrow i} (\omega) =
| \langle X^{N_e-1 N_h-1}, f|\, {\cal {\hat P}} \,| X^{N_e N_h}, i \rangle |^2 \\
\,f_i(T)\; \delta(E^i_{X^{N_e N_h}} - E^f_{X^{N_e-1 N_h-1}} -\hbar \omega).
\end{multline}

\noindent Here ${\cal {\hat P}}$ is the polarization operator, which
destroys an electron and a hole with opposite spin to create a photon with
circular polarization.
The $\delta$-function describing the energy resonance condition is replaced in
practice by a Lorentzian curve with band width $b=0.5$ meV.
We assume thermal equilibrium for the initial states, so that
$f_i(T)$ is the $i$-state Fermi distribution function at temperature $T$.
Unless otherwise stated, only the lowest energy (fundamental) transition is 
studied, as it is the strongest.
This transition involves the ground states of $X^{N_e N_h}$ and $X^{N_e-1 N_h-1}$.
For low temperature, this means that the recombination involves essentially
$s$-shells of the FD spectrum.
Thus, the differences in the behavior of the excitonic complexes we
study do not arise from different symmetries of the occupied orbitals,
but simply from the different correlations in each case.

To compare with the notation in related experiments\cite{DalgarnoPRB},
the energies of the transitions can be written as:

\begin{subequations}
\begin{align}
\label{eq3a}
E_{PL}(X^0)&=E_{X^0}=E^s_e+E^s_h+V_{eh}^{11}, \\
\label{eq3b}
E_{PL}(X^+)&=E_{X^+}-E^s_h=E^s_e+E^s_h+V_{hh}+2\,V_{eh}^{12}, \\
\label{eq3c}
E_{PL}(X^-)&=E_{X^-}-E^s_e=E^s_e+E^s_h+V_{ee}+2\,V_{eh}^{21}.
\end{align}
\end{subequations}

\noindent Here $E_{X^0}$, $E_{X^+}$ and $E_{X^-}$ are the ground state energies
of the neutral, positively charged and negatively charged excitons,
respectively.  $E^s_e$ ($E^s_h$) is the energy of the single-particle $s$
orbitals for electrons (holes), $V_{ee}$ ($V_{hh}$) is the electron-electron (hole-hole) repulsion energy, 
which we calculate as the difference in energy between the two-electron
(two-hole) ground state with and without Coulomb interaction.
Likewise, $V_{eh}^{N_e N_h}$ is the electron-hole attraction energy,
which we calculate as the difference in energy between the excitonic
species with and without including Eq.~(\ref{eq2}) term in the Hamiltonian.
Note that $V_{eh}^{N_e N_h}$ depends on the number of carriers of each type in the QD.


\section{Results}

We consider an InGaAs/GaAs QD of height $2.5$ nm. 
In this material, the electron (hole) effective mass is $m_e^*=0.05$ ($m_h^*=0.45$),
the conduction (valence) band is 350 (200) meV and the dielectric constant is $\epsilon=12.9$.
While this QD structure compares well to Ref.~\onlinecite{DalgarnoPRB} samples, we note
that similar findings are obtained if GaAs/AlGaAs QDs are studied.
The $N_e$ electron and $N_h$ hole correlated states are calculated separately
using a FCI built on all the possible Slater determinants which can be formed from the 
56 lowest-energy FD spin-orbitals.
The excitonic states are then calculated with a basis that includes the Hartree products formed
by several tens of low-energy few-electron and few-hole states.

\subsection{Electron and hole wavefunctions with the same lateral extension}

\begin{figure}[h]
\includegraphics[width=0.4\textwidth]{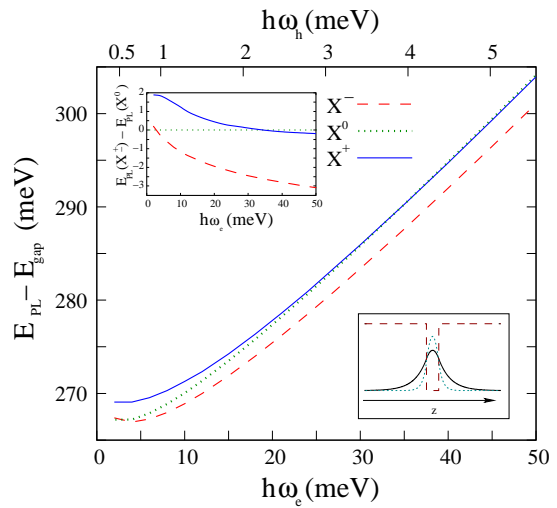}
\caption{(Color online). Recombination energies versus lateral confinement. The electron (hole)
confinement frequency is given in the bottom (top) axis. The characteristic length is set to be
the same for both kinds of particles. Upper-left inset: energy shifts with respect to the
recombination energy of $X^0$.
Lower-right inset: electron (solid line) and hole (dotted line)
single-particle wavefunction in the quantum well (dashed line).}\label{Fig1}
\end{figure}

In a first set of calculations, we vary the confining frequency of the electron
and that of the hole ensuring that the two kinds of particles have the same
confinement length in the plane, $\sqrt{\hbar/m_e^* \omega_{e}}=\sqrt{\hbar/m_h^* \omega_{h}}$.
This is known to be often the case in self-assembled QDs.\cite{HohenesterPSSaPSSb,BesterPRB}
Note that in this way, the single-particle electron and hole wavefunctions are symmetric
in the in-plane direction -the relevant one for the Coulomb-induced
configuration mixing-, and all the differences in the correlation regime can be traced back
to the different level spacing, which is obviously denser for holes.
Electron and hole wavefunctions are still different along the vertical direction,
the hole being more confined due to its larger mass, as shown in the lower inset in Fig.~\ref{Fig1}.
This electron-hole asymmetry, which is present in most kinds of epitaxial QDs,\cite{BesterPRB,KalliakosAPL}
reduces the electron-hole overlap and hence tends to make the electron-hole
attraction ($V_{eh}^{11}$) smaller than electron-electron ($V_{ee}$) and hole-hole ($V_{hh}$) repulsion, 
which avoids artificial cancelations\cite{LelongSSC} of the binding energies of trions obtained when 
$|V_{eh}^{11}|=|V_{ee}|=|V_{hh}|$.

In Fig.~\ref{Fig1} we compare the recombination energies of the excitonic complexes
for variable lateral size of the QD. 
For most confinement regimes, the relative energies agree
with reported measurements in self-organized QDs of different sizes\cite{RodtPRB,DalgarnoPRB} and
early theoretical studies\cite{LelongSSC}: $X^+$ is blueshifted with respect to $X^0$, while $X^-$
is redshifted. However, we also find the possibility of deviations from the usual sequence in the 
strong and weak confinement regime.
This is better seen in the upper inset of the figure, which represents the energy shifts of $X^+$
and $X^-$ with respect to $X^0$:
$X^+$ becomes redshifted for very strong confinement ($\hbar \omega_e >  40$ meV), 
while $X^-$ becomes blueshifted for very weak confinement.

\begin{figure}[h]
\includegraphics[width=0.4\textwidth]{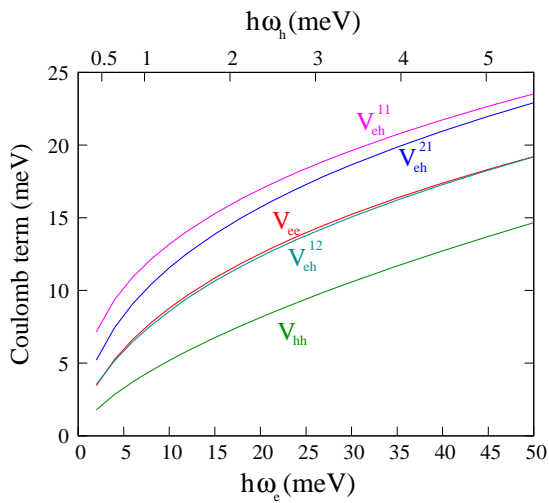}
\caption{(Color online). Absolute value of the Coulomb terms contributing to the exciton
and trion energies of Fig.~\ref{Fig1} (see Eqs.~(\ref{eq3a})-(\ref{eq3c}).}\label{Fig2}
\end{figure}

The result above can be rationalized by comparing the different Coulomb terms contributing to
the recombination energies [see Eqs.~(\ref{eq3a})-(\ref{eq3c})], which are shown in Fig.~\ref{Fig2}.
The redshift of $X^-$ occurs because $E_{PL}(X^-)-E_{PL}(X^0)=V_{ee}-2V_{eh}^{21}+V_{eh}^{11} <0$,
and the blueshift of $X^+$ because $E_{PL}(X^+)-E_{PL}(X^0)=V_{hh}-2V_{eh}^{12}+V_{eh}^{11} > 0$.
It is worth noting a few points regarding the Coulomb terms in the figure:

(i)$V_{ee} > V_{hh}$ for all confinement strengths.
This is in spite of the heavy mass of holes, which should lead to $V_{ee} < V_{hh}$ in
a perturbational approach.\cite{DalgarnoPRB,WarburtonPRB}
This is because, unlike electrons (which have too much kinetic energy),
holes are able to localize well apart in the dot, thus minimizing Coulomb repulsion
(see Fig.~\ref{Fig4} below). In other words, holes are strongly correlated, even in
strongly confined QDs ($\hbar \omega_e=50$ meV).

(ii) $V_{eh}^{11}$ decreases upon the inclusion of an additional carrier ($V_{eh}^{11} > V_{eh}^{21}
> V_{eh}^{12}$).
This is because the Coulomb repulsion separates the two identical carriers,
thus reducing the overlap with the other kind of carrier (which remains in the center of the dot).
At $\hbar \omega_e$=30 meV, the addition of one electron to form $X^-$ reduces
the Coulomb attraction in about 5\%, close to experimental estimates.\cite{DalgarnoPRB}
The addition of one hole, according to our prediction, has a much stronger effect
as the Coulomb attraction is reduced in about 20\%. This is another signature of
the strong hole correlation.

(iii) In general, $V_{eh}^{N_e N_h}$ is larger (in modulus) than $V_{ee}$ and $V_{hh}$.
This is surprising if we recall that the asymmetry of the electron
and hole wavefunctions (in the vertical direction) reduces the electron-hole overlap.
This is again due to electronic correlations, which enable important redistributions of the
charge in the QD through configuration mixing, maximizing attractions ($V_{eh}^{N_e N_h}$)
and minimizing repulsions ($V_{ee}$, $V_{hh}$).
A paradigm is the fact that $|V_{eh}^{11}|$ is clearly larger than $|V_{ee}|$.

The above observations are in qualitative agreement with the data inferred in
Ref.~\onlinecite{DalgarnoPRB} (and extends the predictions towards other confinement regimes).
The only discrepancy is in point (i): while we predict $V_{ee} > V_{hh}$ they
estimate $V_{ee} \sim V_{hh}$.
This may be due to the uncertainty in the confinement potential (for example,
a smoother potential in the $z$ direction would reduce the electron-hole asymmetry)
or in the experimental parameters.
Note however that the values of the Coulomb terms in our calculations are
obtained from the recombination energies of Fig.~\ref{Fig1} 
-which match well the experiment- without further approximations or assumptions. 

\begin{figure}[h]
\includegraphics[width=0.4\textwidth]{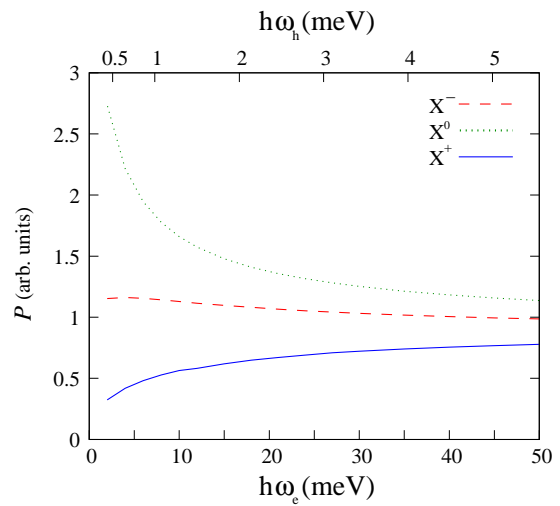}
\caption{(Color online). Recombination probability versus lateral confinement for
the exciton and trion states of Fig.~\ref{Fig1}.}\label{Fig3}
\end{figure}

We next investigate the recombination probability of $X^0$, $X^+$ and $X^-$.
Figure \ref{Fig3} represents the recombination probability for different lateral 
confinement frequency and temperature $T=0$ K.
The recombination probabilities always follow the relation ${\cal P} (X^0) > {\cal P}(X^-) > {\cal P}(X^+)$,
in agreement with the experiment.\cite{DalgarnoPRB,rate}
In particular, in the strong confinement regime the recombination probability functions
are rather flat and obey
${\cal P} (X^0)/{\cal P} (X^+) \approx 1.2$ and ${\cal P} (X^0)/{\cal P} (X^-) \approx 1.7$,
close to the average experimental values of $1.25$ and $1.58$.
As the confinement strength decreases, ${\cal P}(X^0)$ increases very rapidly.
This behaviour is in agreement with previous predictions\cite{BryantPRB,QuePRB,CorniPRB} 
and is due to the increasing electron-hole correlations, 
an effect which is sometimes referred to as exciton superradiance\cite{HanamuraPRB}.
A different behavior is however observed for trions.
${\cal P}(X^-)$ increases only slowly. This is an indication that the repulsive
electron-electron correlation partially compensates for the electron-hole attraction.
On the other hand, ${\cal P}(X^+)$ decreases instead of increasing. This is because
hole-hole correlations dominate over electron-hole ones, and this reduces the
overall electron-hole overlap.

The strong effect of hole correlations upon the charge density shape can be
visualized in Fig.~\ref{Fig4}. The left panel shows the normalized charge
densities of the particles which constitute the $X^+$ complex, 
while the right panel shows that of the $X^-$ one. The relevant result is that, for $X^+$,
the hole with spin down has a dip in the center, where the electron lies.
This is the signature of hole correlations. Clearly, no such signature is
observed for the pair of electrons in $X^-$.
Since the spin down hole is the one that recombines with the spin up electron,
the low PL of $X^+$ reported in Fig.~\ref{Fig3} is readily understood.
This also explains the small value of $V_{hh}$ as compared to $V_{ee}$,
discussed in Fig.~\ref{Fig2}.\\

\begin{figure}[h]
\includegraphics[width=0.4\textwidth]{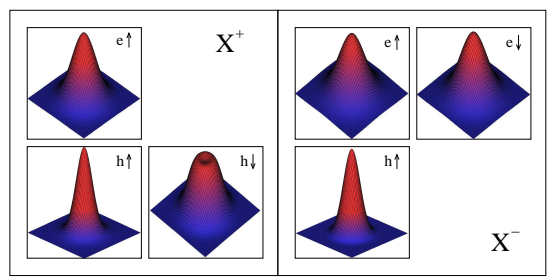}
\caption{(Color online). Charge density of the particles constituting the
trion complexes: $X^+$ (left panel) and $X^-$ (right panel).
The pictures correspond to a QD with $\hbar \omega_e=30$ meV.
e and h stand for electron and hole, and the arrows indicate the spin.
The charge densities are averaged over $z$.
For $X^+$ one of the holes exhibits a dip in the center -the
position of the electron- due to the hole-hole repulsion and the
strong hole correlation. This does not happen for electrons in $X^-$.}\label{Fig4}
\end{figure}

\subsection{Electron and hole wavefunctions with different lateral extension}

So far we have assumed that the single-particle electron and hole wavefunctions have about the same
length in the plane. As mentioned above, this is a good description for usual
self-assembled InGaAs QDs,\cite{BesterPRB} but not necessarily for other kinds of QDs.
This is because electrons and holes feel the band-offset potentials, the strain and
the composition fluctuations in a different way. As a result, hole wavefunctions in
the in-plane direction may be more or less delocalized than electron ones
(the tighter localization in the vertical direction is more robust because the strong
confinement overrides all other effects).\cite{BesterPRB,CadePRB}
In quantum wires this asymmetry has been shown to strongly influence the properties of
trions.\cite{SzafranPRB}
In this section, we investigate its effect on our previous results.

\begin{figure}[h]
\includegraphics[width=0.4\textwidth]{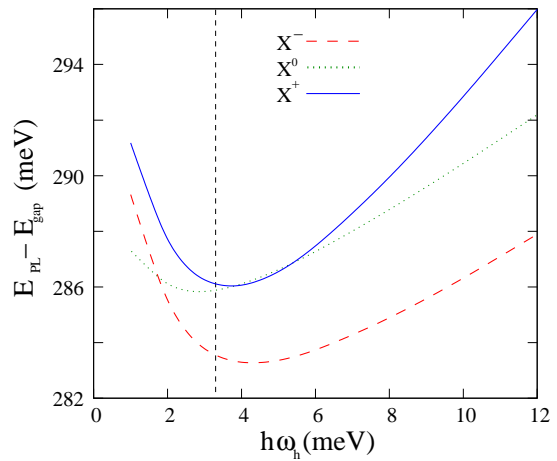}
\caption{(Color online). Recombination energies versus hole lateral confinement
for a fixed electron confinement $\hbar \omega_e=30$ meV. The dashed vertical line
indicates the hole confinement leading to symmetric electron and hole wavefunctions.
To the left (right) of the line, the hole is less (more) localized than the electron.}\label{Fig5}
\end{figure}

We consider a QD with fixed electron confinement $\hbar \omega_e=30$ meV,
which is close to the average value in Ref.~\onlinecite{DalgarnoPRB} samples,
and then vary the hole confinement frequency.
The recombination energies of $X^0$, $X^-$ and $X^+$ in this system are
illustrated in Fig.~\ref{Fig5}.
For $\hbar \omega_h=3.33$ meV (dashed vertical line), the electron and hole
wavefunctions have the same extension. This is the case studied above,
and the result is that of the experiments: $X^+$ is slightly  blueshifted with
respect to $X^0$, while $X^-$ is visibly redshifted.
If we move to the right ($\hbar \omega_h > 3.33$ meV), the hole wavefunction is
more localized than that of the electron. While this barely affects the redshift
of $X^-$, for $\hbar \omega_h \approx 4-5.5$ meV the blueshift of $X^+$ is suppressed.
About 25\% of the QDs investigated in Ref.~\onlinecite{DalgarnoPRB} exhibited
this deviation, which suggests that in such dots holes were slightly more
localized than electrons.
Conversely, if we move to the left ($\hbar \omega_h < 3.33$ meV), the hole wavefunction
is less localized than that of the electron (as in pure InAs QDs\cite{BesterPRB}).
Note that this is a regime of very strong hole correlation.
It may then occur that $X^-$ is redshifted (this was never observed in
Ref.~\onlinecite{DalgarnoPRB} samples).

\begin{figure}[h]
\includegraphics[width=0.4\textwidth]{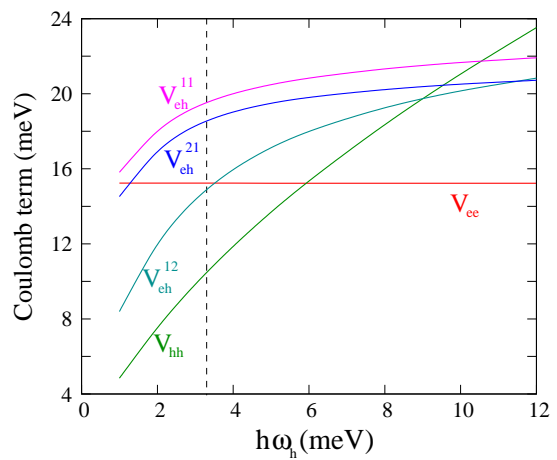}
\caption{(Color online). Absolute value of the Coulomb terms contributing to the
exciton and trion states of Fig.~\ref{Fig5}.
The dashed vertical line indicates the hole confinement where
electron and hole wavefunctions are symmetric.}\label{Fig6}
\end{figure}

The qualitative differences in the recombination energies shown in Fig.~\ref{Fig5}
brought about by the electron-hole asymmetry, and ensuing changes in the regime of
hole confinement, can be again rationalized in terms of the the Coulomb contributions
to Eqs.~(\ref{eq3a})-(\ref{eq3c}). These are plot in Fig.~\ref{Fig6}.
For example, one can see that for very weak hole confinement
$V_{eh}^{11} \sim V_{ee} > V_{eh}^{21}$, which explains the blueshift of $X^-$
($E_{PL}(X^-)-E_{PL}(X^0)=V_{ee}-2V_{eh}^{21}+V_{eh}^{11} > 0$).
In addition, Fig.~\ref{Fig6} gives a clear insight into the effect of hole correlations.
While $V_{ee}$ is obviously insensitive to changes in the hole confinement,
the other terms display a strong, non-trivial dependence. The dependence is especially
strong for $V_{hh}$ and $V_{eh}^{12}$ because they involve two holes.
As we move right from the symmetric electron-hole case (dashed vertical line),
the hole confinement increases.
At about $\hbar \omega_h \approx 6$ meV $V_{hh}$ exceeds $V_{ee}$.
This starts being consistent with a simple perturbational Coulomb picture\cite{WarburtonPRB}
and indicates that hole correlations are decreasing.
Further right, $V_{hh}$ overcomes all the electron-hole terms,
meaning that the hole configuration mixing is no longer able to
maximize (minimize) the attraction (repulsion) terms enough to
compensate for the single-particle electron-hole asymmetry,
which renders $V_{eh}^{N_e N_h}$ smaller than $V_{hh}$.

\begin{figure}[h]
\includegraphics[width=0.4\textwidth]{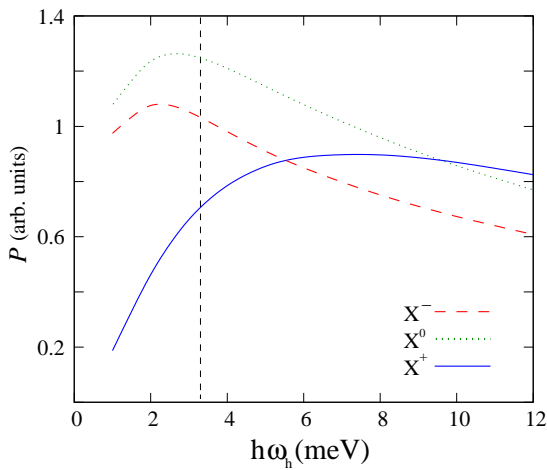}
\caption{(Color online). Recombination probability versus hole lateral confinement for
the exciton and trion states of Fig.~\ref{Fig3}.
The dashed vertical line indicates the hole confinement where
electron and hole wavefunctions are symmetric.}\label{Fig7}
\end{figure}

Fig.~\ref{Fig7} shows the recombination probability of $X^0$, $X^+$ and $X^-$
as a function of the hole confinement frequency. For $X^0$ and $X^-$
the maximal PL is close to point of symmetric
electron and hole wavefunctions (dashed vertical line), where the
single-particle overlap is largest, and it decreases away from this point.
Actually, the maximum is slightly shifted towards the left of the
dashed line because $X^0$ and $X^-$ exhibit superradiant PL
with increased Coulomb correlations (recall Fig.~\ref{Fig3}),
which compensates for a moderate single-particle asymmetry.
The behavior is different for $X^+$ because there is no superradiance
and because the hole-hole repulsion pushes the two holes away from
the electron charge peak, in the center of the QD (as in Fig.~\ref{Fig4}).
As the hole confinement strenght increases, the two holes are squeezed
towards the peak of the electron wavefunction and ${\cal P} (X^+)$ increases.
As a result, the sequence observed in Ref.~\onlinecite{DalgarnoPRB}
(${\cal P} (X^0) > {\cal P}(X^-) > {\cal P}(X^+)$) holds only in the regime
of weak and intermediate hole confinement ($\hbar \omega_h < 6$ meV).
Therefore electrons and holes must have similar lateral extension.
For an electron with $\hbar \omega_e=30$ meV, a symmetryc hole wavefunction
has $\hbar \omega_h=3.33$ meV. If we compare the corresponding two-electron
and two-hole ground states, we find that the dominant configuration in both
cases is the doubly-occupied $s$-shell. However, for electrons the weight of
this configuration is $98.5$ \%, while for holes it is $53.5$ \%.
The latter figure is even smaller than that of electrons in etched QDs
with $\hbar \omega_e \approx 2$ meV,\cite{KalliakosAPL,GarciaPRLKalliakosNP}
which gives an idea of the strength of hole correlations.

Finally we mention that we have run simulations assuming a lighter hole
mass, $m_h^*=0.25$. This value, assuming identical electron and hole
wavefunction in the lateral direction, described well some experiments
with InGaAs QDs.\cite{RaymondPRL}
The same qualitative trends as found here were observed, but the relative
recombination energies did not match those measured in Ref.~\onlinecite{DalgarnoPRB}.
Hole correlations were still important (for $\hbar \omega_e=30$ meV, a symmetric
hole required $\hbar \omega_h=6$ meV, which gives a dominant two-hole configuration
weight of 74\%).

\subsection{Temperature dependence}

\begin{figure}[h]
\includegraphics[width=0.4\textwidth]{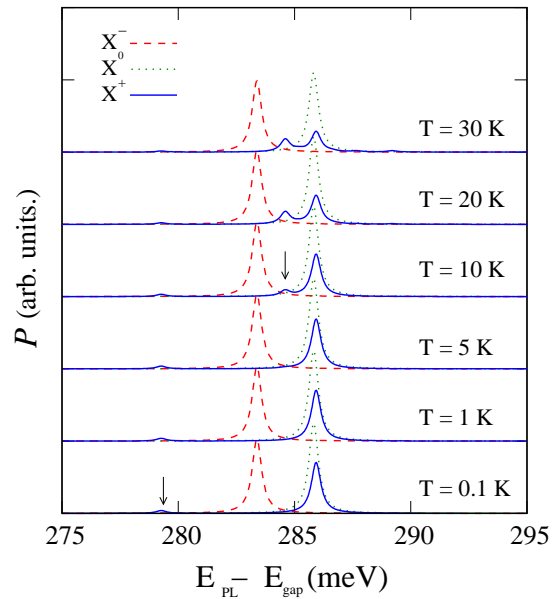}
\caption{(Color online). Emission spectrum of $X^0$, $X^+$ and $X^-$ as a function of the
temperature. The spectra are offset vertically. The arrow at $T=0.1$ K points at the
correlation-induced satellite of $X^+$,
while the arrow at $T=10$ K points at the resonance coming from the first excited state
of $X^+$, which has been activated thermally.}\label{Fig8}
\end{figure}

The markedly different energy structure of $X^+$ as compared to $X^0$ and $X^-$
reflects itself in the calculated temperature dependence of the PL spectra of the 
three excitonic species, shown in Fig.~\ref{Fig8}. 
The plot corresponds to a QD with $\hbar \omega_e=30$ meV and $\hbar \omega_h=3.33$ meV
(symmetric electron and hole wavefunctions).
In this figure, we consider not only the lowest excitonic transition, but also a few 
excited ones which may acquire appreciable population with increasing temperature. 
For the temperatures under study however these transitions play a minor role only.

At the lowest temperature, one can see that $X^+$ alone exhibits a weak satellite
peak (pointed at by an arrow) well below the fundamental transition.
This is due to the strong configuration mixing of holes, which introduces a sizeable
contribution from the $s\,d_z$ configuration to the otherwise pure $s^2$ configuration
of the two-hole ground state.
This mixing enables the recombination of an $s$-shell electron with a $d$-shell hole.
Similar features have been predicted for the excited states of excitons and multi-excitons,
owing to electron-hole correlations,\cite{Pawel_book,HohenesterPSSaPSSb}
as well as for highly charged excitons.\cite{EdigerNP}
Here we show that for positive trions, such features show up for the ground state as well,
this being a signature of the stronger correlation regime.
To our knowledge this resonance has not been explicitly reported in experiments,
but recent high-resolution PL measurements of single InGaAs QDs at
low temperature ($T=5$ K) revealed a number of small satellite peaks below the
fundamental transition of $X^+$ which were however absent for $X^0$ and $X^-$.\cite{EdigerNP}

As the temperature increases, the population of the exciton or trion excited states
increases at the expense of the ground state. For $X^0$ and $X^-$ this has no
observable results up to a few tens of Kelvin, because the lowest-lying excitations are quite far
in energy from the ground state (a few meV). However, for $X^+$ -due to the larger density of states-
moderate temperatures suffice to yield visible changes in the spectrum. This can be seen in
Fig.~\ref{Fig8}. With increasing temperature, the PL of the fundamental
and the satellite peaks of $X^+$ decreases, and a new resonance shows up at $\sim$ 284 meV
(indicated by an arrow at $T=10$ K). This resonance corresponds to the recombination of
an $s$-shell electron and an $s$-shell hole, with a second hole remaining in the $p$ shell.
The different sensitivity to temperature can be used as a means for distinguishing
positively charged excitons from neutral and negative species.
Temperature-dependent PL measurements in individual self-assembled InAs QDs seem 
consistent with this prediction, as they show a fast decrease of the fundamental $X^+$ 
resonance and the appearance a satellite peak right below it 
with increasing temperature (see Fig.~2b in Ref.\onlinecite{CadePRB}).

\section{Conclusion}

Both our accurate effective mass-FCI calculations and recent experimental observations 
of recombination energies and rates point to an important role of configuration mixing of 
valence band holes in the dynamics of trions.
Our results are in quantitative agreement with experimental finding if
electrons and holes are taken to have similar lateral extension.
Under such conditions, hole correlations are clearly non-negligible.
Our results also show that in self-assembled (and weaker confined) QDs
Coulomb correlations lead to $|V_{eh}^{N_e N_h}| > |V_{ee}| > |V_{hh}|$.
More generally, the large increase of recombination probability of $X^0$ as a
function of the dot size -- a well-known result arising from the enhanced electron-hole
Coulomb correlations\cite{BryantPRB,QuePRB,CorniPRB}--, is suppressed almost completely
for $X^-$ and reversed for $X^+$, whose intensity decreases with dot size. 
This is due to the electron-electron and hole-hole Coulomb terms, which compensate (and overcome) 
the excitonic attraction. We also predict that signatures of the distinct hole energy structure
can be found in the specific temperature dependence of the $X^+$ spectrum.

\begin{acknowledgments}
We are grateful to F. Troiani and M. Rontani for helpful discussions.
We acknowledge financial support from the Italian Minister for Research through FIRB-RBIN06JB4C and
PRIN-2006022932, from Cineca Calcolo Parallelo 2007-2008, from the
Marie Curie project MEIF-CT-2006-023797 and a Ramon y Cajal fellowship (J.I.C.).
\end{acknowledgments}


\begin{thebibliography}{46}
\bibitem{rate}
Although the experiment reports recombination rates rather than probabilities, the former are 
directly proportional to the latter (see e.g.~Ref.~\onlinecite{WimmerPRB}), so that the relative 
values are the same.
\end{thebibliography}
\end{document}